\documentclass{aastex}
\usepackage{spr-astr-addons}
\usepackage{url}

\RequirePackage{color}

\newcommand{\angstrom}{\mbox{\normalfont\AA}}

\begin{document}

\title{EXTENDED RED EMISSION: OBSERVATIONAL CONSTRAINTS FOR MODELS}
\slugcomment{Not to appear in Nonlearned J., 45.}
\shorttitle{Short article title}
\shortauthors{Authors et al.}

\author{Adolf N. Witt\altaffilmark{1}} \and \author{Thomas S.-Y. Lai\altaffilmark{1}}

\altaffiltext{1}{Ritter Astrophysical Research Center, University of Toledo,  Toledo, Ohio 43606, USA; corresponding author: awitt@utnet.utoledo.edu}

\begin{abstract}
Extended Red Emission (ERE) is a widely observed optical emission process, present in a wide range of circumstellar and interstellar environments in the Milky Way galaxy as well as other galaxies. Definitive identifications of the ERE carriers and the ERE process are still a matter of debate.
Numerous models have been proposed in recent decades, often developed without consideration of the growing body of observational constraints, which by now invalidate many of these models.
This review focuses on the most well-established observational constraints which should help to delineate the way toward a generally accepted explanation of the ERE and an understanding of its place in the radiation physics of the interstellar medium.

\end{abstract}



\section{Introduction}
Interstellar grains represent about one percent of the mass of the interstellar medium (ISM) in the Milky Way galaxy and other similar stellar systems. However, the role played by interstellar grains in the physics and chemistry of the ISM is disproportionately much larger. Grains are the principal depositories of elements heavier than helium in the ISM \citep{Snow1996}, and they provide the surfaces needed for the dominant process of formation of molecular hydrogen, the most abundant interstellar molecules by far \citep{Cazaux2002}. Even more important is the interaction between photons of the interstellar radiation field and grains through the processes of scattering and absorption. In combination, these two processes are the cause of interstellar extinction, which effectively hides most parts of the Milky Way galaxy from direct view at ultraviolet (UV) and optical wavelengths \citep{Mathis1990}. While scattering is usually the dominant process, leading to diffuse radiation phenomena observed as diffuse galactic light \citep{Leinert1998} and reflection nebulae, absorption of photons by grains has even more important consequences. Absorption of the more energetic UV photons by primarily the smaller grains results in the ejection of electrons via the photoelectric effect, which is the main mechanism for heating the gas in the diffuse ISM (Hollenbach 1989). Much of the energy of absorbed photons leads to heating of interstellar grains, which is the ultimate source of powering the continuous emission at mid- and far infrared and sub-millimeter wavelengths of galaxies. Additionally, some fraction of the absorbed photon energy is used to power fluorescence or photoluminescence (PL) observable at optical wavelengths such as blue luminescence \citep[BL:][]{Vijh2004, Vijh2005} and extended red emission (ERE; \citealt{Witt2004}). 

While ERE and BL are often mentioned in conjunction with each other, they do not appear to share the same carriers and certainly not the same excitation/emission process. Both phenomena were first seen in the Red Rectangle (RR) nebula (Fig. \ref{fig:RR}). However, their spatial distributions within the nebula are drastically different \citep{Vijh2006}. The RR nebula has long been known as a strong source of the unidentified infrared emission (UIE) features, now generally attributed to polycyclic aromatic hydrocarbon (PAH) molecules. The measured BL intensity distribution in the Red Rectangle follows closely the spatial distribution of the 3.3 $\mu m$ UIE band, the emission dominated by the smallest PAH molecules. The spectrum of the BL, with a sharp peak near a wavelength of 380 nm and a red tail extending to about 500 nm, closely matches that of the fluorescence spectra of small neutral PAHs with three to four aromatic rings such as anthracene and pyrene \citep{Vijh2004}. In a discussion of the sizes of PAH molecules, \citet{Allamandola1989}, based on observed ratios of the 3.3 $\mu m$ and 11.3 $\mu m$ PAH bands, identified the RR nebula as an object distinguished by smaller PAHs (10 to 50 carbon atoms), in contrast to other nebulae where PAH sizes were found in the size range of 35 to 120 carbon atoms. The spatial distribution of the BL in the RR nebula is consistent with environments where such molecules would be expected to be predominantly neutral \citep{Vijh2006}. Thus, the existence of neutral PAH molecules such as anthracene and pyrene was, in essence, predicted by the mid-infrared UIE data. The discovery of a fluorescence spectrum in the form of the BL was, therefore, an important but not unexpected confirmation. There have not been serious challenges to the identification of the BL carriers as neutral PAH molecules with 14 to 16 carbon molecules. This is in strong contrast to the situation with regard to the ERE, where many competing carrier proposals have been advanced over the years (Table \ref{tab:ere_models}). This review, therefore, deals with the observational constraints these models must be subjected to for a generally accepted explanation of the ERE to emerge.

\section{The discovery of ERE}
\begin{figure}[t]
 	\includegraphics[width=\columnwidth]{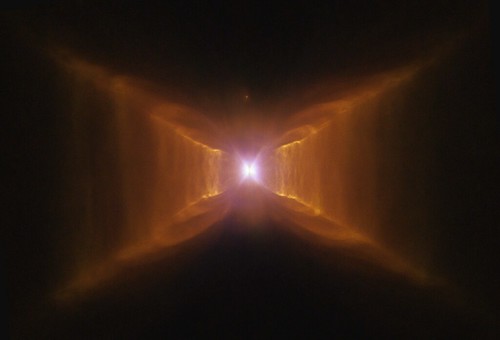}
    \caption{The Red Rectangle nebula, produced by mass outflows from the central star HD 44179. The outflows occur within two cones in the horizontal direction, constrained by a massive disk readily visible in this image taken with the Hubble Space Telescope. The field of view of this image is about 20 by 25 arcseconds. This image is a result of reprocessing of openly available images from the NASA/ESA Hubble Space Telescope by Judy Schmidt.}
    \label{fig:RR}
\end{figure}

ERE was observed first in a peculiar bi-polar nebula associated with the post-asymptotic giant branch (post-AGB) star HD 44179 \citep{Cohen1975, Schmidt1980}, introduced as the Red Rectangle (RR) nebula earlier. (Figure \ref{fig:RR}). The bi-polar shape of the RR nebula is the result of mass-outflows from an accretion disk surrounding a low-mass companion star of HD 44179. The accretion disk is being supplied with mass and accretion-powered energy by rapid mass overflow (rate $\sim$ 2--5 $\times$ 10$^{-5}$ M$_{\odot}$/yr) from the post-AGB primary \citep{Witt2009}. ERE, seen in spectra of the RR nebula as a broad emission band, extending in wavelength from about 5400 $\angstrom$ to about 7600 $\angstrom$, with a peak near 6600 $\angstrom$, is strongly concentrated near the walls of the outflow cones \citep{Vijh2006}. This gives the RR nebula its characteristic near-rectangular shape in low-resolution images taken at red wavelengths, while its morphology is roughly circular at blue wavelengths \citep{Vijh2006}, where the dominant emission mechanism is scattering by dust. While a fair number of bi-polar nebulae associated with other post-AGB stars are known, the RR nebula is the only one exhibiting ERE. We attribute this to the fact that only in the RR nebula is the accretion disk sufficiently massive and hot to produce the far-UV photons needed for the excitation of the ERE as well as for the ionization of a small H II region surrounding the central binary system. The discovery of ERE in the RR nebula was largely a consequence of the high ERE intensity relative to dust-scattered light, due to the fact that the dust scattering in the RR nebula occurs under large scattering angles near 90 degrees, which is comparatively inefficient. By contrast, ERE is emitted isotropically and is thus independent of the nebular scattering geometry.

Initially, ERE was considered a peculiar phenomenon unique to the then still enigmatic RR nebula, with no far-ranging implications for the interstellar medium in general. The dust in the RR nebula is most likely the result of local condensation of heavy elements in the mass outflow from the post-AGB star HD 44179, resulting in a potentially unique family of grains and molecules. This changed with the introduction of CCD detectors into observational astronomy, which provided the capability of greatly enhanced detection sensitivity at red and near-infrared wavelengths. Taking advantage of the fact that the broad ERE band almost perfectly coincides with the photometric R-band allowed the detection of ERE in numerous reflection nebulae by means of imaging photometry \citep{Witt1984,Witt1985,Witt1986}, which was soon followed by spectroscopic confirmations \citep{Witt1988, Witt1989a, Witt1990}. Ultimately, ERE was found in most reflection nebulae illuminated by stars earlier than spectral type A0, corresponding to an effective temperature $>$ 10,000 K \citep{Darbon1999}.

Given that in most instances reflection nebulae are regions where ordinary interstellar gas and dust is concentrated at higher densities with stronger radiation fields from recently formed stars, the dust characteristics there should be similar to those in the general diffuse ISM. This led to the suggestion that ERE should be associated with the general diffuse ISM as well. This was confirmed by detections of ERE in galactic cirrus clouds \citep{Guhathakurta1989, Guhathakurta1994, Szomoru1998, Witt2008} and in the general high-latitude diffuse ISM \citep{Gordon1998}. These observations indicate that the ERE intensity integrated over the photometric R-band is comparable to that of the intensity of the diffuse galactic light in the R-band, the latter resulting from scattering of the interstellar radiation field by dust in interstellar space.

It is worth noting that ERE was also found in regions with very high radiation densities such as the bar in the Orion nebula \citep{Perrin1992}, the H II region NGC 7635 \citep{Sivan1993}, the halo of the starburst galaxy M82 \citep{Perrin1995}, the 30 Doradus nebula in the LMC \citep{Darbon1998}, and the galactic compact H II region Sh152 \citep{Darbon2000}. Such observations are complicated by the fact that ERE is observed superimposed on a complex spectrum composed of dust-scattered light, atomic recombination continuum, and numerous strong emission lines, not just simply dust-scattered light as in reflection nebulae. A common characteristic of the ERE band in high-radiation-density environments is a significant shift of the peak wavelengths of the respective ERE bands toward longer wavelengths, well beyond 700 nm and even beyond 800 nm.

A set of ERE detections with particular significance were those in planetary nebulae \citep{Furton1990, Furton1992}, because dust and molecules in planetary nebulae are the result of local formation in the expanding material making up the nebulae. Many planetary can be classified as carbon-rich or oxygen-rich, depending on whether the C/O ratio in the nebular gas is either $>$ 1 or $<$ 1.
The dust and molecules forming in these respective environments are then assumed to reflect the respective chemistry, being dominated either by carbonaceous or oxide-related structures, respectively. In a large sample with roughly similar numbers of carbon- and oxygen-rich planetary nebulae, ERE was detected only in carbon-rich objects. This suggests that the ERE carrier is most likely carbonaceous in nature.

The observational detections of ERE relied on a variety of observational techniques, which were reviewed in detail by \citet{Witt2004}. We refer the reader to the individual papers for further details.

\section{Observational constraints for ERE carrier models}

The ERE literature abounds with proposals for the likely carriers for the ERE phenomenon \citep{Witt2004, Lai2017}, as shown, together with their seminal references, in Table 1. There are several reasons for this. Observations of ERE are challenging and observational constraints emerged only gradually over time. Most model papers are based on laboratory experiments conducted under conditions quite different from those under which ERE occurs in astrophysical settings, with a principal aim of the approximate reproduction of a particular ERE spectrum. It seems highly unlikely that the multitude of models can all be correct. Rather, it should be our aim to identify one or more processes and carrier models that satisfy all well-established observational constraints. We will briefly review these constraints in the following sections.

\begin{table*}
	\centering
	\caption{ERE models}
	\begin{tabular}{ll} 
		\rule{0pt}{4ex}
		\textbf{I. Classical Fluorescence/Photoluminescence} & References \\
		\hline
		\hline
		Hydrogenated Amorphous Carbon & \citet{Duley85}; \citet{Witt1988};\\
		& \citet{Furton1993}; \citet{Mulas2004}; \\
		& \citet{Godard10}\\
		Polycyclic Aromatic Hydrocarbons & \citet{D'Hendecourt86}\\
		Quenched Carbonaceous Composite	& \citet{Sakata92}; \citet{Wada2008}\\
		Fullerence C$_{60}$ & \citet{Webster93}\\
		Biofluorescence & Hoyle \& Wickramasinghe (\citeyear{Hoyle96},\citeyear{Hoyle99})\\
		Silicon Nanoparticles & \citet{Ledoux98}; \citet{Witt98}\\
		& \citet{Zubko1999} \\
		Carbon Clusters & \citet{Seahra99}\\
		Nanodiamonds & \citet{Chang06}; \citet{Lu2017}\\
		MgSiO$_3$ Silicate & \citet{Thompson13}\\
		Graphene Oxide Nanoparticles & \citet{Sarre2019} \\
		\hline
		\rule{0pt}{4ex}
		\textbf{II. Two-Step Processes}\\
		\hline
		\hline
		PAH Di-cations & \citet{Witt2006}\\
		PAH Dimer Cations & \citet{Rhee2007}\\
		\hline
		\rule{0pt}{4ex} 
		\textbf{III. Recurrent Fluorescence in Isolated Molecules}\\
		\hline
		\hline
		Poincare Fluorescence & \citet{Leger1988}\\
		Thermally Excited Molecules & \citet{Duley2009}\\
		\hline
	\end{tabular}
	\label{tab:ere_models}
\end{table*}

\subsection{Spectral characteristics}

ERE is seen as a broad emission band, starting in wavelength near 540 nm. The peak wavelength of the ERE band varies from object to object and even within a given object in the sense that it shifts to larger values as the offset from the illuminating source is reduced. The critical factor appears to be the density of the illuminating radiation field \citep{Smith2002}. In the diffuse ISM with the lowest radiation field density, the ERE peak is seen near 600 nm \citep{Szomoru1998}, while reflection nebulae with average radiation fields about two to three orders of magnitude stronger exhibit peaks in the range from 650 nm to 700 nm \citep{Witt1990}. In H II regions, where the radiation density is still higher by another one or two orders of magnitude, the peak wavelength of the ERE band shifts into the range from 750 nm to 800 nm and beyond \citep{Perrin1992, Sivan1993}. The shift in peak wavelength toward larger values is accompanied by a reduction in the photon conversion efficiency, measured by the ratio of the number of ERE photons per absorbed UV-optical photon in the 91.2 to 540 nm range, by about two order of magnitude \citep{Smith2002}.

At the low spectral resolutions at which ERE spectra are typically observed, the ERE band does not exhibit any recognizable spectral structure. One possible exception is the RR nebula, where a set of sharp, narrow emission bands appear superimposed on the ERE band \citep{Schmidt1991, vanWinckel2002, Wehres2011}. These sharp bands are spatially closely correlated with the ERE in the RR nebula \citep{Schmidt1991}, but they have not been observed in any other well observed ERE spectrum. Furthermore, the only other location where these narrow emission bands have been seen is the circum-stellar dust region of the R Corona Borealis star V854 Cen \citep{Rao1993, Oostrum2018}, however with the distinct absence of the broad ERE band. It appears most likely, therefore, that the sharp emission bands seen in the RR nebula and in V 854 Cen are a phenomenon separate from ERE \citep{Lai2020}, with a different set of carriers but similar excitation requirements.

\subsection{Excitation of the ERE}
The ERE is a luminescence process whose intensity is clearly a function of the UV-optical radiation density in the environments in which the ERE is observed \citep{Smith2002}. As ERE is typically found in regions with neutral rather than ionized gas, the spectrum of the illuminating radiation extends from 91.2 nm in the far- UV into the optical and near-infrared wavelength ranges. In objects with ionizing radiation such as H II regions or planetary nebulae, the ERE is seen on the surfaces of clumps of molecular gas, where absorption by atomic hydrogen has screened out the ionizing Lyman continuum. There, ERE often appears in the form of narrow filaments \citep{Witt1989b, Vijh2006, Witt2006, Lai2017}, resulting from projection effects when such surfaces are seen approximately edge-on. 

The first question any proposal for a possible source of ERE must face is whether a spectrum with the characteristics described in Sect. 3.1 can be produced under illumination conditions that include the full UV-optical spectrum as the source of excitation. Many luminescent materials will produce a photo-luminescence (PL) spectrum resembling the ERE, when excited with a blue laser source at a wavelength just short of 500 nm. However, there is strong circumstantial evidence that suggests that it is far-UV radiation rather than optical light that is the cause of ERE.

First evidence for the dominant role of UV photons in the production of ERE was published by \citet{Witt1985}. This was based on the rapid decrease of the ERE intensity with increasing optical depth in extended nebulae, consistent with the much stronger internal absorption within the nebulae at UV wavelengths. This was confirmed with much stronger observational evidence by \citet{Witt2006} in the case of the ERE in the reflection nebula NGC 7023 and by \citet{Lai2017} for the externally illuminated nebula IC 63. The latter studies identified photons within the energy range 10.5 eV $<$ E $<$ 13.6 eV as the most likely sources of ERE production. An analysis of ERE observations in a wide range of nebulae by \citet{Darbon1999} provided powerful confirmation of the exclusive role of UV photons in the production of the ERE by showing that only nebulae illuminated by stars with T$_{eff}$ $\simeq$ 10,000 K or higher exhibit ERE, while nebulae illuminated by cooler stars (T$_{eff}$ $<$ 7000 K) do not. It is noteworthy that stars with T$_{eff}$  $>$  10,000 K emit photons in the energy range 10.5 eV $<$ E $<$ 13.6 eV, while stars cooler than that do not.

The dominant role of far-UV photons in the production of ERE at the exclusion of optical photons of lesser energy essentially eliminates ERE models relying on classical fluorescence/PL processes with excitation at optical wavelengths. It does not, however, exclude models involving two-step processes, where the first step uses far - UV photons to prepare the ERE carriers, e.g. through double-ionization of large PAH molecules, which might then be excited to fluoresce after excitation by optical photons \citep{Witt2006}. Finally, the dominant role of far-UV photons in ERE excitation is an essential element of the recurrent fluorescence model, also known as Poincare fluorescence \citep{Leger1988}, in which marginally stable molecular ions absorb far-UV photons, followed by emission of low-energy optical photons via the process of inverse internal conversion \citep{Nitzan1979, Duley2009}. Excellent experimental confirmation of the latter process has been established in recent years \citep{Martin2013, Chandrasekaran2014, Ebara2016, Ito2014}. It should be noted that recurrent fluorescence is a potentially powerful mechanism by which marginally stable molecules or molecular ions in interstellar space can significantly improve their chances of survival in environments with high UV radiation densities, by being able to shed a large fraction of their vibrational energy through multiple electronic de-excitations occurring on a much faster timescale than vibrational transitions.

\subsection{Photon conversion efficiency}
The photon conversion efficiency is one of the most important characteristics of any PL process. It is measured by the ratio of the number of photons emitted in the emission band and the number of photons in the excitation spectrum that are absorbed by the PL carrier. Therefore, for an estimate of the photon conversion efficiency, both the band-integrated PL intensity and the spectrum and intensity of the illuminating radiation field need to be known. In the case of the ERE, this became possible for the first time with the detection of the ERE in the diffuse ISM \citep{Gordon1998}, because the density and spectrum of the interstellar radiation field (ISRF) responsible for the illumination are quite well known (Mathis et al. 1983; Porter \& Strong 2005). The band-integrated ERE intensity at high galactic latitudes was found to be about 7 $\times$ 10$^{-6}$ erg/cm$^2$ s sr, approximately equal to the intensity of the dust-scattered diffuse galactic light in the R-band in the same directions. Assuming that all photons of the ISRF integrated over the wavelength range from 540 nm, the short-wavelength edge of the ERE band, to 91.2 nm, the Lyman discontinuity of atomic hydrogen, contribute equally to the excitation of the ERE, Gordon et al. (1998) estimated a photon conversion efficiency of (10 $\pm$ 3)\% for the process producing ERE in the diffuse ISM. We note that this is a hard lower limit to the photon conversion efficiency. As discussed in the previous section, only photons in the far-UV portion of the spectrum appear to be responsible for the excitation of the ERE. This greatly reduces the number of photons available for excitation, leading to a corresponding increase in the photon conversion efficiency. \citet{Witt2006} found that the number of photons in the energy range 10.5 eV $<$ E $<$ 13.6 eV absorbed per unit time in the diffuse ISM is smaller than the number of ERE photons emitted per unit time. This implies that even with an assumed photon conversion efficiency of 100\%, classical PL processes cannot account for the production of the ERE. This suggests that either two-step processes must be invoked or photon conversion efficiencies of more that 100\% need to be considered. Note that the conversion efficiency is defined in terms of the number of photons involved in the conversion rather than in terms of their energy.
Recurrent fluorescence \citep{Leger1988, Duley2009} in suitably small, marginally stable molecules or molecular ions ($\sim$18 to 36 carbon atoms) are expected to produce at least three ERE photons upon absorption of a single far-UV photon, which would readily satisfy this constraint. However, the photon conversion efficiency of recurrent fluorescence decreases rapidly as the size of the molecules increases, while the energy of the exciting photons in the diffuse ISM is limited by the Lyman absorption edge (E = 13.6 eV) of atomic hydrogen. Upon absorption of a 13.6 eV photon, a larger molecule will not reach a sufficiently high vibrational temperature to permit the excitation of low-lying electronic states via inverse internal conversion.

In contrast to the case of the diffuse ISM, the ERE photon conversion efficiency in nebulae illuminated by hot stars, when evaluated in the same manner as applied by \citet{Gordon1998}, was found to be lower by one to two orders of magnitude compared to the efficiency in the diffuse ISM \citep{Smith2002}. Even though the ERE in nebulae can be comparable in intensity to that of the scattered light underlying the ERE band \citep[e.g.][]{Lai2017} as is the case in the diffuse ISM, the exciting spectrum in typical nebulae rises steeply in intensity from the optical to the far-UV by about two orders of magnitude, while the exciting ISRF spectrum in the diffuse ISM is comparatively flat in the UV-optical wavelength range. The reason that ERE is found to be brighter in nebulae than in the diffuse ISM is simply due to the much higher radiation density of the exciting photons, not as a result of a high quantum yield. 

Any ERE PL model needs to explain how the proposed carriers are modified under the enhanced radiation conditions found in nebulae to simultaneous produce ERE spectra in which the wavelength of peak emission shifts to longer wavelengths with increasing density of the prevailing far-UV radiation field while exhibiting a decreasing photon conversion efficiency at the same time.

\subsection{ERE in photo-dissociation regions}
In contrast to the diffusely scattered light in nebulae, ERE is observed most prominently in narrow filamentary structures on the surfaces of molecular clouds directly illuminated by the central star(s). In the RR nebula, the ERE appears mainly on the inner surfaces of bi-polar outflow cones, which are in a direct line of sight to the central source of the nebula \citep{Schmidt1991, Vijh2006}. In projection on the plane of the sky, this results in the well-known X-shaped structure of the RR nebula when observed in red light. In NGC 2023, the first nebula in which ERE was found other the RR nebula, the ERE appeared to be widely distributed throughout the western and southern portions of the nebula \citep{Witt1984}. Closer examination revealed that the ERE was strongly concentrated in multiple filaments in NGC 2023, coinciding with illuminated surfaces on molecular clumps viewed edge-on \citep{Witt1989b}. Today, such filamentary structures are recognized as photo-dissociation regions (PDRs), where the gaseous medium transitions from a low-density neutral or even ionized phase to a much denser molecular phase under the influence of the far-UV radiation from a nearby massive star over a short distance range. 

\begin{figure}[t]
 	\includegraphics[width=\columnwidth]{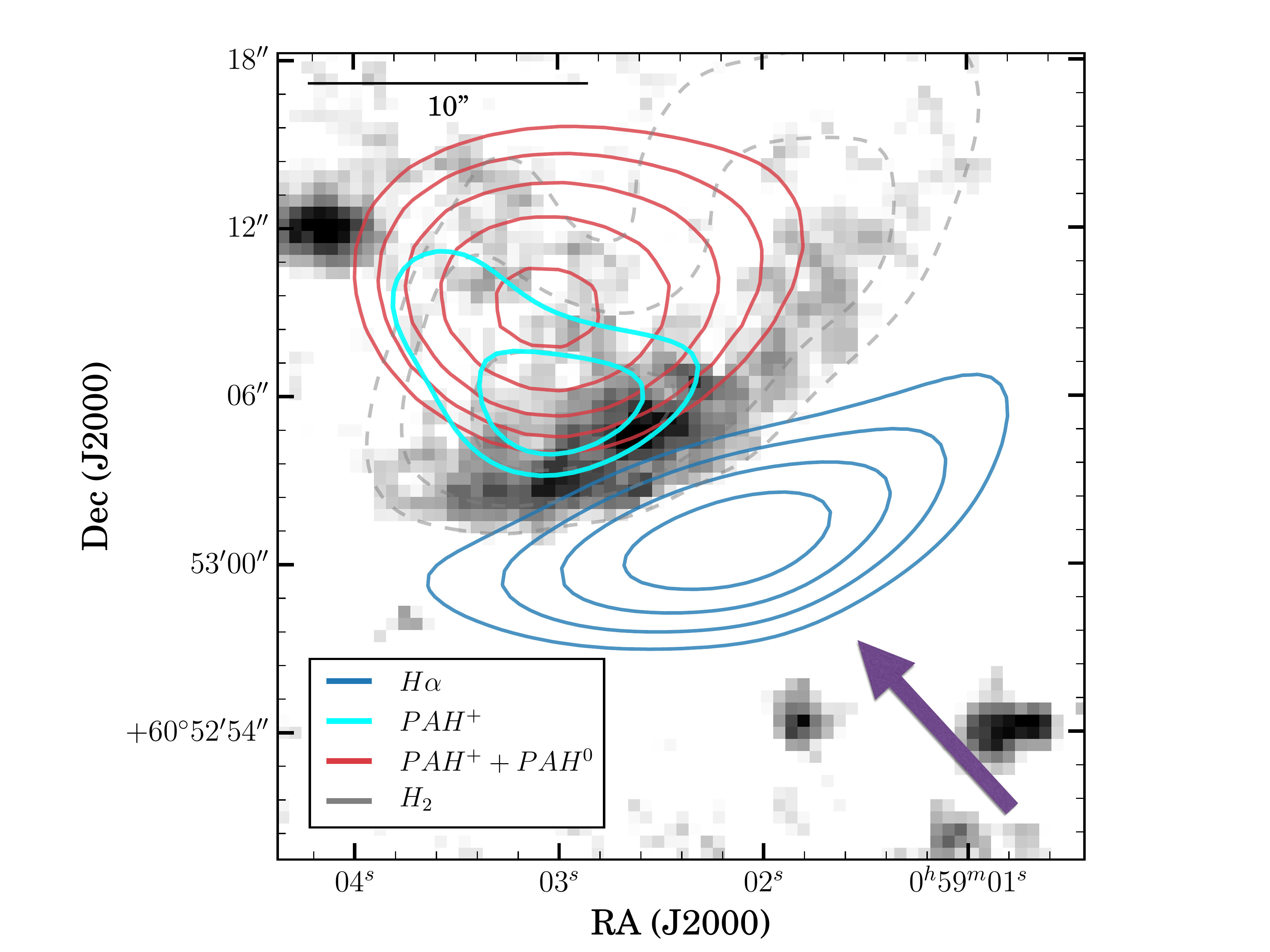}
    \caption{The map in grayscale shows one of the two brightest ERE filaments in IC 63. The arrow indicates the direction of the incoming radiation. The ERE filaments are located immediately behind the hydrogen ionization front traced by H$\alpha$ emission (blue) and in front of the ionized PAHs (cyan). ERE resides in the front ends of the peak emission of neutral PAHs (red) and pure H$_{2}$ rotational lines (dashed gray).}
    \label{fig:layering}
\end{figure}

When viewed edge-on, the emission structure and composition of a PDR is determined by the depth to which radiations of different wavelengths are able to penetrate into the molecular environment. The angular width of the PDR as seen in a given emission process is inversely proportional to a product of distance to the observer, the density, and the opacity for the radiation needed to excite the process. When applied to the ERE filaments seen in the reflection nebula NGC 7023, \citet{Witt2006} found that the ERE filaments are nearly identical in structure and width as corresponding filaments seen in UV-pumped near-IR 1 - 0 S(1) H$_2$ emission. The latter process is excited by far-UV photons absorbed in the Lyman and Werner bands of H$_2$ in the wavelength region 110.4 to 91.2 nm. This suggests that the ERE carriers receive their primary excitation in the same wavelength region; if longer wavelength were involved, the width of ERE filaments would be greater than that of the H$_2$ filaments. This finding was confirmed with a study of the ERE in the PDR of the emission/reflection nebula IC 63, which is illuminated externally by the B 0.5 IV pe star Gamma Cas \citep{Lai2017}. This PDR, being closer and less dense than the structures in NGC 7023, is more easily resolved and shows a clear sequence of emission regions with increasing distance from the source: ionized gas (H-alpha), ERE, UV-pumped H$_2$ rotational line emission, ionized polycyclic aromatic hydrocarbons (PAHs), and neutral PAHs (Figure \ref{fig:layering}). The most intense ERE coincides with the narrow atomic hydrogen region of the PDR where molecular hydrogen is almost fully dissociated and where exciting radiation in the 110.4 to 91.2 nm range is not attenuated by the opacity of molecular hydrogen. These findings are fully consistent with the empirical data of \citet{Darbon1999} and other early studies of ERE excitation.

\subsection{Possible ERE-DIB connection}
One of the first and still most promising models for the ERE process was the idea of recurrent fluorescence by marginally stable, highly isolated molecules \citep{Leger1988}. To produce a broad emission band with the spectral characteristics of the ERE requires the presence of a large family of molecules and molecular ions with low-lying electronic states (E $<$ 2.3 eV) that can survive in a wide range of radiation environments with photon energies as high as E = 13.6 eV. As noted by \citet{Witt2014}, such a family of molecular structures may indeed exist in the still mostly unidentified molecular carriers of diffuse interstellar bands (DIBs). \citet{Witt2014} noted that the frequency distribution of DIB transitions in wavelength space is remarkably similar to the shape of the ERE spectrum in the diffuse ISM and in reflection nebulae, with most of the $>$ 500 DIBs so far discovered \citep{Hobbs2008, Hobbs2009} falling into the wavelength range from 600 nm to 700 nm. Such low-lying electronic states with high transition probabilities to the ground state are in fact a requirement for efficient recurrent fluorescence. In turn, recurrent fluorescence provides an efficient and fast mechanism for shedding a large fraction of energy such molecules receive when absorbing far-UV photons. This greatly enhances the survival probability against photo-fragmentation for such molecules, allowing them to build up sufficiently large column densities for producing detectable absorption features in transitions from the ground to the low-lying electronic states.

To test this possible synergy between the DIB and ERE processes, \citet{Lai2020} observed the spectrum of a faint (V = 18.24) background star, IC 63 Star \#46, with a line of sight that penetrates the brightest ERE filament in the PDR of the nebula IC 63. About two thirds of the total cumulative reddening of the star’s flux (E(B-V) = 1.23) occurs within the PDR while the remainder is caused by dust in the background of IC 63. \citet{Lai2020} detected the five strongest DIBs expected to be detectable in a low-resolution, low S/N spectrum: DIB4428, DIB5780, DIB5797, DIB6284, and DIB6614. The strongest and widest of these bands, DIB4428, exhibits an equivalent width of about twice the value expected for normal diffuse ISM lines of sight with identical reddening, while the other four DIBs show approximately normal equivalent widths for the total cumulative amount of reddening. This suggests that the extreme radiation environment in the PDR does not affect the survival of the carriers of these five DIBs, when the far-UV radiation density is several hundred times that present in the normal diffuse ISM. Moreover, the carriers of the strong DIB4428 appear to be present with a significantly enhanced abundance, possibly as a result of top-down production by photo processing of carbonaceous nanoparticles in the PDR. While these findings do not prove a direct connection between the ERE and DIB phenomena, they are entirely consistent with the proposed synergy. Clearly, more observations of this type are needed. If supported by further work, a demonstrated connection between ERE and DIBs would greatly constrain the range of models for the ERE that would be considered viable.

\section{Implications for ERE models}
Interestingly, there appears to be a far greater number of models in the literature for the ERE than for other unexplained interstellar phenomena. This reflects two facts: the abundance of PL processes and substances capable of producing ERE-like spectra under certain illumination conditions and the challenges and time requirements associated with the ERE observations that led to significant observational constraints, which could have limited the range of models. Also, the models are generally not mutually exclusive, although each model purports to provide the full explanation. It is not the aim of this review to discuss individual models in detail; such discussions can be found in the review by \cite{Witt2004} and in \citet{Lai2017}, but we will provide some guidelines for possible progress.

Following \citet{Lai2017}, we can divide the range of published ERE models into three categories; classical PL models, two-step processes, and recurrent fluorescence by isolated molecules and molecular ions. Most classical PL models are based on reports on laboratory experiments where ERE-like luminescence or fluorescence spectra were observed from substances illuminated at wavelengths just short of the ERE band. Most classical PL processes are most efficient, when the Stokes shift, the energy difference between the exciting radiation and the resulting PL is relatively small, of the order of 0.2 eV. Also, no PL photons can appear at wavelengths shorter than that of the exciting radiation. When exciting radiation at ultraviolet wavelengths is applied, much closer to the illumination conditions prevailing in actual ERE sources, the resulting PL either displays much wider spectra, extending well into the green and blue spectral ranges where no ERE has been observed, or the conversion into red PL photons occurs with a greatly reduced efficiency. The ERE process needs to be highly efficient; otherwise PL models encounter new constraints set by the cosmic abundances of the needed materials in interstellar space. An example of such a limitation was presented by \citet{Zubko1999} for the case of silicon nanoparticles, a popular ERE model at the time. While silicon has a comparatively high cosmic abundance, most of it is depleted in the form of silicates, making up the larger fraction of the mass of interstellar grains. The main reasons for the failure of most classical PL models for the ERE is that they ignore the observational constraints on ERE excitation under interstellar conditions and that they do not provide evidence for the required high photon conversion efficiency.

Two-step processes \citep{Witt2006, Rhee2007} are based on the idea that far-UV photons are necessary to create the ERE carrier, not necessarily excite the ERE itself.
In these cases, doubly ionized PAH molecules or singly ionized PAH clusters, respectively, must be maintained in their states of ionization by far-UV photons but can be excited to fluoresce in the ERE band by absorbing the far more abundant photons throughout the UV-optical wavelength range short of the ERE band. Observational evidence for the ubiquitous presence of PAH-like structures is abundant, and the models respect the fact that ERE appears only where the needed ionizing photons are present. The principal shortfall of these models is the lack of experimental evidence, showing that they do indeed fluoresce in the spectral range of the ERE band with a sufficiently high quantum yield.

Recurrent fluorescence (RF), also known as Poincare fluorescence, to explain ERE was first invoked by \citet{Leger1988}, following the theoretical exploration of the process of inverse internal conversion by \citet{Nitzan1978, Nitzan1979}. RF occurs in highly isolated molecules, more likely molecular ions, which are vibrationally highly excited as a result of absorbing far-UV photons. If such structures possess low-lying electronic states, inverse internal conversion allows vibration energy residing in the electronic ground state to electronically excite such low-lying states, followed by electronic transitions to ground. The latter result in the emergence of fluorescence photons. Only in recent years has experimental confirmation of this process been forthcoming \citep{Martin2013, Chandrasekaran2014, Ebara2016, Ito2014}.
In order to produce an ERE spectrum with its observed wavelength coverage, the RF model \citep{Leger1988, Duley2009} requires a large family of molecular structures with low-lying states covering this wavelength range. As pointed out by \citet{Witt2014}, the carriers of the over 500 diffuse interstellar bands, widely thought to be molecular in nature and observed ubiquitously observed throughout interstellar space, do in fact satisfy this requirement. As shown by \citet{Leger1988}, if the molecules are sufficiently small (18 - 36 carbon atoms), the photon conversion efficiency of the RF process can exceed 300\%, with the highest efficiency resulting from the absorption of photons near the Lyman limit of atomic hydrogen with E = 13.6 eV. It is remarkable that the RF model automatically satisfied the requirements of far-UV excitation and extremely high photon conversion efficiencies many years before these observational constraints were well established. The recent observations of the strong presence of DIB carriers within regions of high ERE intensity by \citet{Lai2020} lends further support to the RF model for the ERE.

At this stage, two-step luminescence processes and the RF model provide the most promising paths to explaining the ERE. However, more experimental data will be required before a conclusive solution can be identified. If the RF process continues to gain support, the likelihood of simultaneously solving two outstanding problems in the ISM, ERE and DIBs, becomes a real possibility.

\acknowledgments        \\            
We gratefully acknowledge 
Judy Schmidt for allowing us to reproduce the image of the Red Rectangle nebula shown in Figure \ref{fig:RR}. The image is a result of her reprocessing of available images from the NASA/ESA Hubble Space Telescope.  https://www.flickr.com/photos/g \\
eckzilla/10159700064/
We also wish to thank the anonymous referee for constructive suggestions which aided in making this review more complete.




\nocite{*}
\bibliographystyle{spr-mp-nameyear-cnd}
\bibliography{ERE_review}

\end{document}